# Temperature Dependence of Volatile Current shoot-up in PrMnO₃ based Selector-less RRAM

S. Lashkare*, A. Bhat*, and U. Ganguly

*Abstract*—**PrMnO₃ (PMO) based Resistance Random Access Memory (RRAM) has recently been considered for selector-less RRAM and neuromorphic computing applications by utilizing its current shoot-up. This current shoot-up in the PMO device is attributed to the thermal runaway in the device. Hence, the understanding of the ambient temperature dependence on the current shoot-up of the PMO device is essential for the various applications that utilize the negative differential resistance (NDR). In this paper, we characterize the ambient thermal dependence of dc IV, accompanied by the development of analytical modeling. First, the temperature dependent current-voltage characteristic and shift in the threshold voltage of the PMO device are shown experimentally. Second, a Joule heating based thermal feedback model coupled with current transport by space charge limited current (SCLC) is developed to explain the experimentally observed NDR region. Finally, the model successfully predicts device behavior over a range of experimental ambient temperatures. As an alternative to TCAD, such a compact and accurate dc model sets up a platform to enable understanding, design with device and systems level simulations of memory and neuromorphic applications.**

## I. Introduction

The cross-bar Resistance Random Access Memory (RRAM) array implementation requires selector devices to reduce sneak path current leakage [1]-[3]. However, the integration of RRAM and selector device in an array is challenging. To avoid this issue, selector-less memory devices which have high non-linearity (current shoot-up in low resistance region (LRS)) are extensively explored [4]-[7]. Recently, PrMnO₃ (PMO) RRAM device is shown to be as selector-less RRAM by utilizing its current shoot-up in LRS [8]-[9]. PMO RRAM is attractive as it is forming less, area scalable with high endurance and retention properties [9]-[10].

Further, PMO RRAM's have been demonstrated as a Rectified Linear Unit (ReLU) type neuron [11] and an oscillatory neuron [12], both of which have applications in neuromorphic computing. In the ReLU type neuron, the PMO devices current shoot-up is used to mimic as the integration function in the neuron [11]. In oscillator neuron, the PMO device is used in series with a capacitor to provide self-sustained oscillations. The oscillation window is defined by highly non-linear hysteretic IV characteristic of PMO device [12]. All these applications utilize the current shoot-up in the PMO device.

The work is partially funded by DST Nano Mission and Ministry of Electronics and IT (MeitY). It was performed at IIT Bombay Nanofab Facility. S. Lashkare is supported jointly by the Visvesvaraya PhD Scheme of MeitY, Government of India, being implemented by Digital India Corporation and Intel Ph.D. Fellowship. S. Lashkare and A. Bhat contributed equally to this work.

S. Lashkare and U. Ganguly are with the Department of Electrical Engineering, Indian Institute of Technology Bombay, Mumbai, 400076, India. (e-mail: sandipl@ee.iitb.ac.in; udayan@ee.iitb.ac.in). A. Bhat was with Department of Electrical Engineering, Indian Institute of Technology Bombay, Mumbai, 400076, India. He is now with University of Michigan, Ann Arbor, USA (ashwinbhat111095@gmail.com)

While many applications have been demonstrated experimentally as discussed above, the mechanisms behind the PMO devices current shoot-up have not been investigated. The dependence of dc IV characteristics on ambient temperature is an important aspect of device behavior, which has not been explored yet. Here, we investigate the temperature dependence on the IV characteristics of PMO devices accompanied by the development of analytical modeling, which is essential for various applications.

## II. Device Fabrication and Characterization

The PMO RRAM devices were fabricated on 4" Si substrate. The $SiO_2$ (300nm) layer is epitaxially grown by thermally oxidizing the Si substrate in wet oxidation furnace. A Ti (20 nm)/Pt (70 nm) layer is then deposited by sputtering. Pt acts as a bottom contact. Then, a PMO layer (65nm) is deposited by RF sputtering. The stack is then annealed at 750°C in $N_2:O_2$ (95%:5%) ambient for 30s to crystallize the PMO layer. Finally, tungsten (W) top contact pads (5$\mu m$ × 5$\mu m$) were created by the photolithography and lift-off process. The device schematic is shown in Fig. 1(a).

All the dc IV characterizations have been done on Agilent B1500A Semiconductor Analyzer.

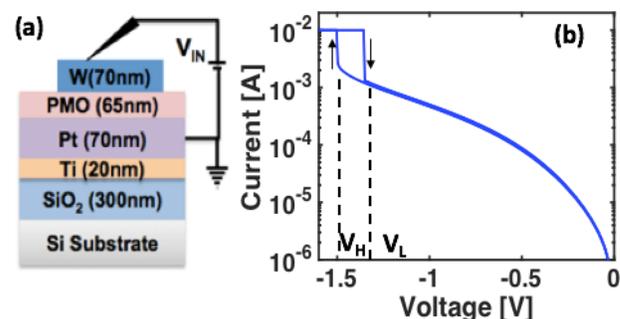

Fig. 1 **Experimental** (a) Device schematic showing PMO sandwiched between W and Pt electrodes. (b) Hysteretic dc IV of PMO device showing a sharp rise and fall in current after a voltage threshold $V_H$ and $V_L$.

## III. Temperature Dependence

The PMO device is semiconducting [13] and highly thermally insulating in nature [14]. The dc IV (voltage as input and current as output) of the PMO device is shown in Fig. 1(b). The device shows a volatile hysteretic IV on double voltage sweep ($0 \rightarrow -1.6V \rightarrow 0V$). After a threshold voltage ($V_H$), the current rises sharply and hits the compliance level. The compliance level is set to avoid the breakdown of the device. The device returns back to its initial state in the reverse voltage sweep ($-1.6V \rightarrow 0$) with a hysteresis. This sharp switching in the IV has been attributed to the self-heating in the device (positive feedback of current and temperature) [15] due to the low thermal conductivity ($0.005\ Wcm^{-1}K^{-1}$) of PMO which makes it >300× thermally resistive than silicon ($1.48\ Wcm^{-1}K^{-1}$) [14]. To study this sharp region, a current sweep (current as input and voltage as output) is performed (Fig. 2(a)). The current controlled Negative Differential Resistance (NDR) region (i.e. reduction in voltage for an increase in current) can be observed in the current sweep. As the temperature is dependent on heat generation (due to Joule heating) and heat–sink (i.e., ambient temperature), the device's ambient temperature is varied and measurements are performed (Fig. 2(b)). It is observed that, as the temperature is increased, the threshold voltage ($V_H$) for sharp rise decreases as the required current for shoot-up is reached at a lower voltage.

bias beyond a threshold voltage ($V_H$), the self-heating enhanced SCLC current flows through the device leading to thermal runaway.

$$J_{Ohmic} = q\mu N_v e^{-\frac{q\phi}{kT}}\left(\frac{V}{L}\right) \quad (1)$$

$$J_{SCLC} = \mu \left(\frac{N_v}{N_T}\right) e^{\left(-\frac{qE_{trap}}{kT_{Device}}\right)} \left(\frac{V^2}{L^3}\right) \quad (2)$$

where, $\mu$= Mobility, $\phi$= Barrier Height, $N_v$= Effective density of states, $E_{trap}$= Trap energy level, $N_T$= Trap density, L = PMO thickness.

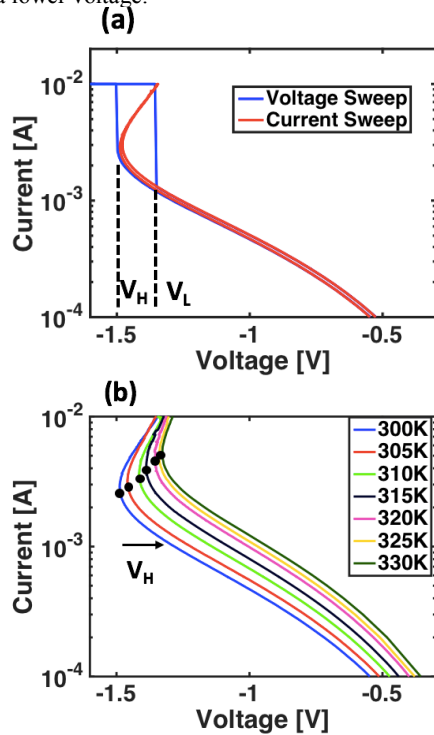

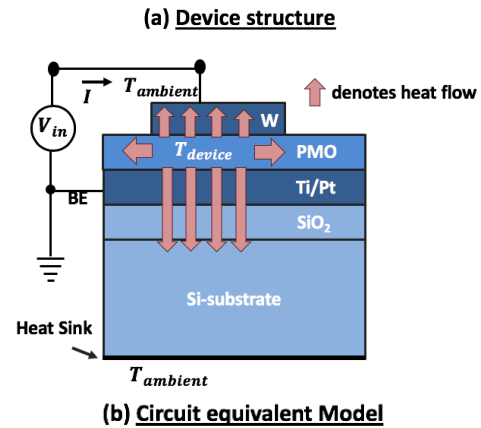

Fig. 2 **Experimental** (a) Voltage sweep showing the sharp rise while the current sweep showing the current controlled negative differential region (NDR) region at room temperature, (b) Current sweep with different ambient temperatures shows a reduction in threshold voltage ($V_H$).

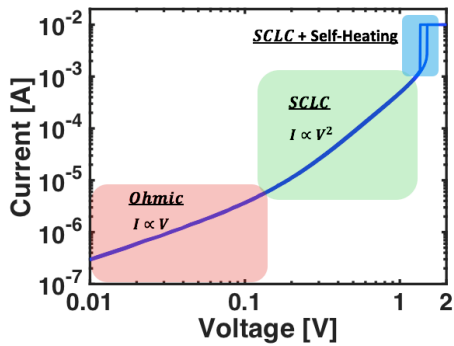

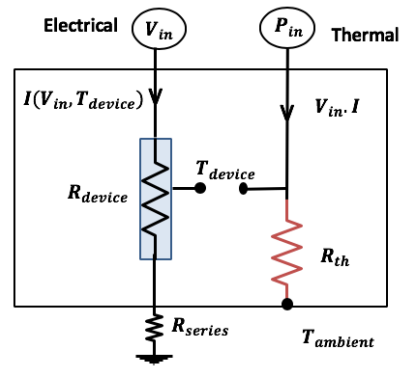

Fig. 3 Log-log plot showing different regions of operation of the device. At low bias, the current conduction is ohmic. With an increase in the bias, SCLC current starts flowing through the device. After a threshold voltage, self-heating enhanced SCLC current flows device giving a sharp current rise.

IV. THERMAL FEEDBACK MODEL

To study the effect of thermal insulation on the IV characteristics of the PMO device, a thermal feedback model is adopted from [16]. The model is modified with the SCLC conduction mechanism as described below.

The PMO devices log-log IV characteristics show different regimes of operation of the device. At low bias, current conduction is ohmic (1). As the bias is increased, it transitions into the space charge limited current (SCLC) regime (2) [17]. On further increasing the

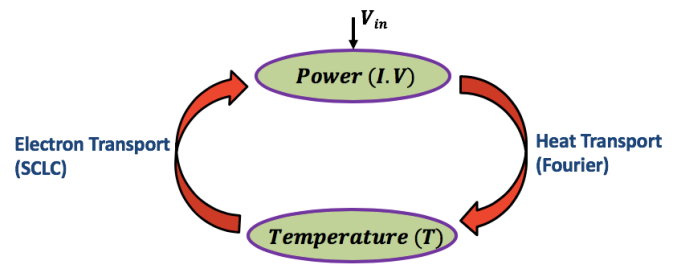

Fig. 4 **Thermal Feedback Model** (a) PMO device schematic showing the path for heat loss in the device, (b) circuit equivalent model showing the current and temperature interdependence, (c) Thermal feedback model incorporating trap-based SCLC current conduction mechanisms to capture the current shoot-up in the device.

The device structure showing the current and heat flow is shown in Fig. 4(a). As the PMO device is highly thermally insulating, the heat is retained within the device leading to an increase in device temperature. An equivalent circuit model shows that the SCLC current is dependent on the applied voltage and device temperature i.e. $I(V_{in}, T_{device})$. The $T_{device}$ is further dependent on input power i.e. $T_{device}(V_{in}, I)$, thereby creating a positive feedback loop (Fig. 4(b)) between current and temperature. A physical feedback model is developed for this positive feedback to solve for the electrical and thermal effects simultaneously to capture the current shoot-up in the PMO device (Fig. 4(c)). The parameters used for the simulations are given in Table I.

**Table I**
Parameters used in model

| Parameter | Value |
|---|---|
| Area | 10 µm² |
| Thickness($t_{PMO}$) | 65 nm |
| Barrier Height(φ) | 0.05 eV |
| Thermal conductivity(k) | $0.5 \frac{W}{mK}$ [14] |
| Mobility (µ) | $3 \frac{cm^2}{Vs}$ [18] |
| Perimitivity ($\epsilon_r$) | 30 [19] |
| Trap Density ($N_t$) | $1.5 \times 10^{20} cm^{-3}$ |

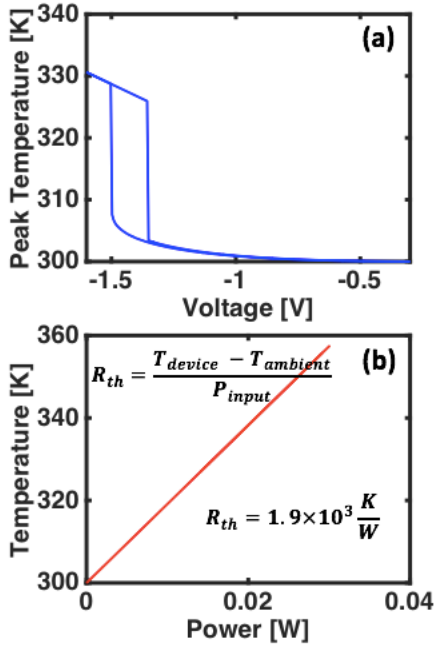

Fig. 5 (a) Calculated temperature by solving 1-D Fourier heat equation, (b) Lumped thermal model for heat transfer to calculate thermal resistance ($R_{th}$) using DCIV (Fig. 4(a)) [16].

### A. Device Parameter Extractions

*1) Thermal Resistance ($R_{th}$)*: The device temperature (Fig. 5(a)) is calculated by solving a steady state Fourier Heat equation given below

$$-k \frac{d^2 T}{dx^2} + c_v \frac{dT}{dt} = \frac{IV}{volume} \quad (3)$$

Where, $k$ = Thermal conductivity, $c_v$ = Specific heat, $volume = A.L = 10 \times 10 \mu m^2 \times 65 nm$

Further, to calculate the equivalent thermal resistance ($R_{th}$) of whole device structure a lumped thermal model is used (Fig. 5(b)) [16]. This lumped thermal model for heat transfer considers entire device to be at the same temperature and is described by

$$R_{th} = \frac{T_{Device} - T_{ambient}}{I.V_{Device}} \quad (4)$$

where, $T_{ambient}$ = ambient temperature, $V_{Device}$ is the voltage drop across the device as opposed to an applied voltage

$$V_{in} = V_{Device} + I.R_{series} \quad (5)$$

The $R_{Series}$ is the parasitic resistance due to probe to device pad contact resistance. $R_{series}$ is assumed to be constant at all temperature ranges for simplicity.

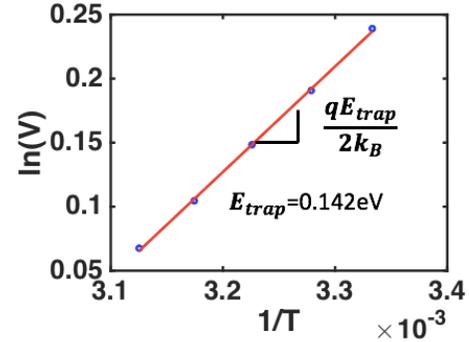

Fig. 6 Trap energy level ($E_{trap}$) extraction from SCLC current by fixing $J_{SCLC}$ to get V for different T in (2).

*2) Trap Energy Level ($E_{trap}$)*: The trap energy is required to calculate the SCLC current (2). $E_{trap}$ is calculated by fixing the SCLC current and then finding the voltage over different ambient temperatures from Fig. 2(b). By rearranging the (2), the $E_{trap}$ can be found as the slope of $\ln(V)$ $vs.$ $1/T$ (Fig. 6) as shown below:

$$\ln(V) = \frac{qE_{trap}}{2k}\left(\frac{1}{T}\right) + c \quad (6)$$

where c is a constant at fixed current.

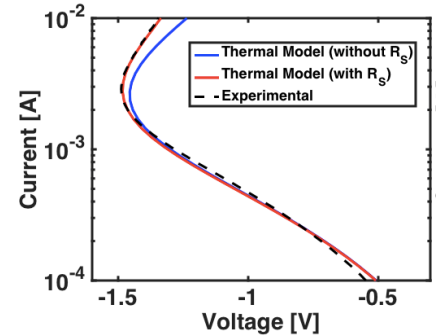

Fig. 7 Excellent matching of current sweep with experiments validates the thermal model by using series resistance of 10Ω.

### V. MODEL VALIDATION WITH EXPERIMENTAL RESULTS

The calculated $R_{th}$ parameter and extracted $E_{trap}$ are then used in the thermal feedback model to calculate $I(V, T)$. Initially, the model did not include the series resistance (10Ω). The series resistance is on the order of probe to device pad contact resistance [18]. Hence, the simulated current (blue) did not show the matching with the experimental results (Fig. 7). However, after addition of the series resistance (red), excellent matching is seen. This matching of simulated current sweep using model with experiments shows the validation of the developed thermal feedback model.

Finally, the model is used to predict the behavior of the devices at different ambient temperatures (Fig. 8). The excellent agreement between the simulation results and experimental measurements at various ambient temperatures further provides strong validation of the proposed modeling methodology.

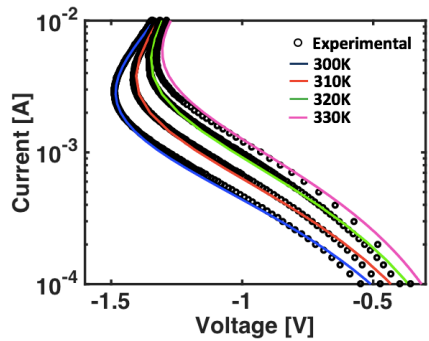

Fig. 8 The model predicting the device behavior with different ambient temperatures.

## VI. Conclusions

In this paper, first, NDR is characterized at different ambient temperatures experimentally. Second, a phenomenological model is developed where current and heat transport are simultaneously solved. The thermal resistance ($R_{th}$) is calculated using a lumped thermal model. The trap energy ($E_{trap}$) is uniquely extracted for various temperatures. These parameters lead to excellent agreement of model vis a vis experiments across a range of temperatures (300-330K). Thus, the study presents the ambient temperature dependence of the dc IV characteristics of PMO based RRAM including the NDR. The observations are quantitatively reproduced by a model of couple electronic current and heat transport over the range of temperature. The analytical model is highly efficient compared to detailed TCAD simulations. Such a study provides an initial platform for understanding and design at the devices and systems level for the various applications of PMO based RRAM in selector-less memory and neuromorphic applications.